\documentclass[twocolumn,showpacs,preprintnumbers,amsmath,amssymb,superscriptaddress]{revtex4}
\usepackage{mathrsfs}

\usepackage{amsmath,amssymb,graphicx,bm}

\begin{document}


\title{Geometrical diagnostic for purely kinetic k-essence dark energy}

\author{Xiang-Ting, Gao}
 \affiliation{College of Physical Science and Technology, Hebei University, Baoding 071002, China}

\author{Rong-Jia, Yang}
\email{yangrj08@gmail.com} \affiliation{College of Physical Science
and Technology, Hebei University, Baoding 071002, China}

\date{\today}

\begin{abstract}
Geometrical diagnostic, involving the statefinder $\{r,s\}$ and
$Om(x)$, is widely used to discriminate different dark energy
models. We apply the statefinder $\{r,s\}$ and $Om(x)$ to purely
kinetic k-essence dark energy model with Dirac-Born-Infeld-like Lagrangian which can be considered as scalar field realizations
of Chaplygin gas. We plot the evolution
trajectories of this model in the statefinder parameter-planes and
$Om(x)$ parameter-plane. We find that the statefinder $\{r,s\}$ and
$Om(x)$ fail to distinguish  purely kinetic k-essence model from
$\Lambda$CDM model at $68.3\%$ confidence level for $z\ll 1$.
\end{abstract}


\maketitle

\section{Introduction}

In the last decade a convergence of independent cosmological
observations suggested that the Universe is experiencing accelerated
expansion. An unknown energy component, dubbed as dark energy, is
proposed to explain this acceleration. Dark energy almost equally
distributes in the Universe, and its pressure is negative. The simplest and most theoretically appealing candidate of
dark energy is the vacuum energy (or the cosmological constant
$\Lambda$) with a constant equation of state (EoS) parameter $w=-1$.
This scenario is in general agreement with the current astronomical
observations, but has difficulties to reconcile the small
observational value of dark energy density with estimates from
quantum field theories; this is the cosmological constant problem. It is thus natural to pursue alternative possibilities to explain
the mystery of dark energy. Over the past decade numerous dark energy models have been proposed,
such as quintessence, phantom, k-essence, tachyon, (Generalized) Chaplygin Gas, DGP, etc. k-essence, a simple approach toward constructing a model for an accelerated expansion of the Universe, is to work with the idea that the unknown dark energy component is due exclusively
to a minimally coupled scalar field $\phi$ with non-canonical kinetic energy which results in the negative pressure \cite{Arm00}. This scenario has received much attention, considerable efforts have been made in understanding the role of
k-essence on the dynamics of the Universe. A feature of k-essence models is that the
negative pressure results from the non-linear kinetic energy of the
scalar field. Secondly, because of the dynamical attractor behavior,
cosmic evolution is insensitive to initial conditions in k-essence
theories. Thirdly, k-essence changes its speed of evolution in
dynamic response to changes in the background equation-of-state.
Here we only concentrate on a special class of k-essence with
Dirac-Born-Infeld-like Lagrangian $p(X)=-V_0\sqrt{1-2X}$. This class of k-essence can be considered as scalar field realizations
of Chaplygin gas \cite{Chim04, Frolov, Kame01, Gorini05} and have been studied intensively (see e. g. \cite{Ahn, Keresztes, Chiba, Yang09, yang, yang08}).

Since more and more dark energy models have been constructed, the
problem of discriminating between various dark energy models is
important. To solve this problem, Sahni et al.\cite{V} and Alam et
al.\cite{U} introduced a geometrical diagnostic, called statefinder.
The statefinder probes the expansion dynamics of the Universe
through higher derivatives of the expansion factor $\dddot{a}$ and
is a natural companion to the deceleration parameter $q$ which
dependent on $\ddot{a}$. The statefinder pair $\{r,s \}$ is defined
as
\begin{eqnarray}
r\equiv\frac{\dddot{a}}{aH^{3}}, ~~~~~s\equiv \frac{r-1}{3(q-1/2)},
\end{eqnarray}
where $a$ is the scale factor, $H\equiv \dot{a}/a$ is the Hubble
parameter and $q\equiv \ddot{a}/(aH^2)$ is the deceleration
parameter.

Trajectories in the $r-s$ plane corresponding to different
cosmological models exhibit qualitatively different behaviors. The
spatially flat $\Lambda$CDM scenario corresponds to a fixed point
$\{r,s\}=\{1,0\}$ in the diagram. Departure of a given dark energy
model from this fixed point provides a good way of establishing the
distance of the model from $\Lambda$CDM. If this distant can be
measured, models can be distinguished. It has been demonstrated that
the statefinder can successfully differentiate between a wide
variety of dark energy models, such as $\Lambda$CDM, quintessence
\cite{V, Sirichai,Linder}, holographic dark energy model
\cite{Granda,Zhang}, Ricci Dark Energy model \cite{Feng}, DGP
\cite{Grigoris}, Generalized Chaplygin Gas Model
\cite{Gorini,Writambhara,Li}, Agegraphic Dark Energy Models
\cite{Wei}, quintom dark energy model \cite{Wu} etc.

Another diagnostic $Om(x)$ was introduced to differentiate
$\Lambda$CDM from other dark energy models except of $\{r,s \}$
\cite{V2}. $Om(x)$ is a combination of the Hubble parameter and the
cosmological redshift and provides a null test of dark energy being
a cosmological constant $\Lambda$. Namely, if the value of $Om(x)$
is the same at different redshift, then dark energy is $\Lambda$
exactly. The slope of $Om(x)$ can distinguish dynamical dark energy
from the cosmological constant in a robust manner both with and
without reference to the value of the matter density, which can be a
significant source of the uncertainty for cosmological
reconstruction. It has been shown that the $Om(x)$ can successfully
differentiate between a wide variety of dark energy models, such as
quintessence\cite{V2}, phantom\cite{V2}, Ricci Dark Energy model
\cite{Feng}, holographic dark energy \cite{Granda,Zhang}, etc.

In this Letter we apply the statefinder $\{r,s\}$ and $Om(x)$ to
purely kinetic k-essence dark energy models with
Dirac-Born-Infeld-like Lagrangian $p(X)=-V_0\sqrt{1-2X}$ which can be considered as scalar field realizations
of Chaplygin gas. We plot the evolution
trajectories of this model in the statefinder parameter-planes and
$Om(x)$ parameter-plane. We find that the statefinder $\{r,s\}$ and
$Om(x)$ fail to distinguish purely kinetic k-essence model from
$\Lambda$CDM model.

In section II, we will briefly review purely kinetic k-essence model.
In section III, we plot the evolutionary trajectories of this model in the statefinder parameter planes.
In the last  section we will give same conclusions.

\section{Briefly Review on k-essence}
As a candidate of dark energy,
k-essence \cite{Arm00} is defined as a scalar field $\phi$ with
non-linear kinetic terms which appear generically in the effective
action in string and supergravity theories, and its action minimally
coupled with gravity generically may be expressed as
\begin{eqnarray}
S_{\phi}=\int d^4x\sqrt{-g}\left[-\frac{R}{2}+p(\phi,X)\right],
\end{eqnarray}
where $X=\frac{1}{2}\partial_{\mu}\phi\partial^{\mu}\phi$, and we
take $8\pi G=1$ throughout this Letter. The Lagrangian $p$ and the
energy density of k-essence take the forms: $p_{\rm k}=V(\phi)F(X)$
and $\rho_{\rm k}=V(\phi)[2XF_{X}-F]$. Here $F(X)$ is a function of
the kinetic energy $X$ and $F_{X}\equiv dF/dX$. The corresponding
equation of state (EoS) parameter and the effective sound speed are
given by
\begin{eqnarray}
\label{3}w_{\rm k}&=&\frac{F}{2XF_{X}-F}, \\
\label{4}c^{2}_{\rm s}&=&\frac{\partial p_{\rm k}/\partial
X}{\partial\rho_{\rm k}/\partial X}=\frac{F_{X}}{F_{X}+2XF_{XX}},
\end{eqnarray}
with $F_{XX}\equiv d^{2}F/dX^{2}$. The definition of the sound speed
comes from the equation describing the evolution of linear
perturbations in a k-essence dominated Universe \cite{Gar99}. In
this Letter, we consider a class of k-essence with constant potential
\cite{Chim04,yang}: $p_{\rm k}(X)=-V_0\sqrt{1-2X}$, where $V_0$ is a
constant. Such models can be considered as scalar field realizations
of Chaplygin gas \cite{Chim04,Frolov, Kame01, Gorini05}. For this
Lagrangian, the EoS parameter and the sound speed take the form
respectively \cite{Yang09},
\begin{eqnarray}
\label{20}w_{\rm k}=-c^{2}_{\rm s}=-\frac{1}{1+2k_0^2a^{-6}},
\end{eqnarray}
where $k_0=\sqrt{2}F_0/2V_0$ is a constant ($-\infty<k_0<+\infty$,
but because of the exponent 2, the case $k_0\geqslant 0$ and the
case $k_0\leqslant 0$ are equivalent). For $k_0=0$, the above EoS
reduces to $-1$; meaning the $\Lambda$CDM model is contained in
k-essence model as one special case. The behavior of the EoS
(\ref{20}), being $\simeq-0$ in the early Universe, runs closely to
$-1$ in the future for $k_0\neq 0$. Such behavior can, to a certain
degree, solve the fine-turning problem \cite{Arm00, Chi}.

\section{Geometrical diagnostic for kinetic k-essence dark energy}
Statefinder $\{r,s\}$ introduced in Refs. \cite{V, U} is a useful method to differentiate kinds of dark energy models.
Researches have shown that it can differentiate $\Lambda$CDM from many dark energy models including $\Lambda$CDM, quintessence
\cite{V, Sirichai,Linder}, holographic dark energy model
\cite{Granda,Zhang}, Ricci Dark Energy model \cite{Feng}, DGP
\cite{Grigoris}, Generalized Chaplygin Gas Model
\cite{Gorini,Writambhara,Li}, Agegraphic Dark Energy Models
\cite{Wei}, quintom dark energy model \cite{Wu}, etc.
For model of dark energy with equation of state $w_{\rm D}$ and
density parameter $\Omega_{\rm D}$, the statefinder parameters
$\{r,s\}$ can be expressed as follows \cite{V}
\begin{eqnarray}
r&=&1+\frac{9}{2}\Omega_{\rm D}w_{\rm D}(1+w_{\rm D})-\frac{3}{2}\Omega_{\rm D}\frac{\dot{w}_{\rm D}}{H},\\
s&=&1+w_{\rm D}-\frac{1}{3}\frac{\dot{w}_{\rm D}}{Hw_{\rm D}}.
\end{eqnarray}

For the purely kinetic k-essence dark energy, the statefinder
parameters and the deceleration parameter can be expressed as
\begin{eqnarray}
r&=&1+9\Omega_{\rm k} \frac{k_{0}^2(1+z)^6}{[1+2k_{0}^2(1+z)^6]^2},\\
s&=&-\frac{2k_{0}^2(1+z)^6}{1+2k_{0}^2(1+z)^6},
\end{eqnarray}
and
\begin{eqnarray}
q=\frac{\Omega_{\rm
m0}(1+z)^3+\frac{2k_{0}^2(1+z)^6-2}{1+2k_{0}^2(1+z)^6}(1-\Omega_{\rm
m0})f(z)}{2\Omega_{\rm m0}(1+z)^3+2(1-\Omega_{\rm m0})f(z)},
\end{eqnarray}
here $f(z)=\exp[3\int_{0}^{z}\frac{1+w_{\rm
k}(z^{'})}{1+z^{'}}dz^{'}]$, and $\Omega_{\rm
k}(z)=\frac{\Omega_{\rm k0}f(z)}{\Omega_{\rm k0}f(z)+\Omega_{\rm
m0}(1+z)^3}$.

Constrained form 307 SNIa data \cite{Kowalski}, the shift parameter
$R$ \cite{wang}, and the acoustic scale $l_{\rm a}$ \cite{wang}, the
best-fit values of the parameters at $68\%$ confidence level were
found to be: $\Omega_{\rm m0}=0.36\pm0.01$ and $k_0=0.067\pm0.011$
\cite{Yang09}. With those parameters, we plot the evolution
trajectories of purely kinetic k-essence model in the statefinder
parameter-planes.

\begin{figure}
\includegraphics[width=8cm]{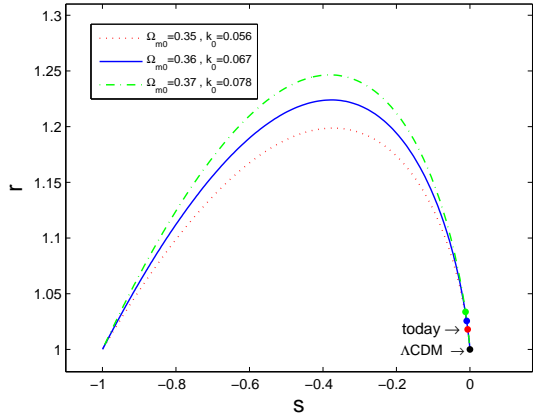}
\caption{Evolution trajectory in the statefinder $s-r$ plane for purely kinetic
k-essence model. The locations of today's point are plotted. The black
dot is $\Lambda$CDM fixed point (0,1).}
\end{figure}

In Fig. 1, $\Lambda$CDM scenario corresponds to a fixed point:
$\{r,s\}=\{1,0\}$. As $s$ varies in the interval $[-1,0]$, $r$ first
increases from $r=1$ to its maximum values and then decreases to the
$\Lambda$CDM fixed point. We clearly see that the `distance' from
today's values of purely kinetic k-essence to $\Lambda$CDM model is
hardly identified in this diagram at $68\%$ confidence level.
Meanwhile today's values get closer and closer to the $\Lambda$CDM
model by decreasing $k_{0}$ and $\Omega_{\rm m0}$. Hence, the
statefinder diagnostic can't discriminate purely kinetic k-essence
model and $\Lambda$CDM model at $68\%$ confidence level for $z\ll
1$.

\begin{figure}
\includegraphics[width=8cm]{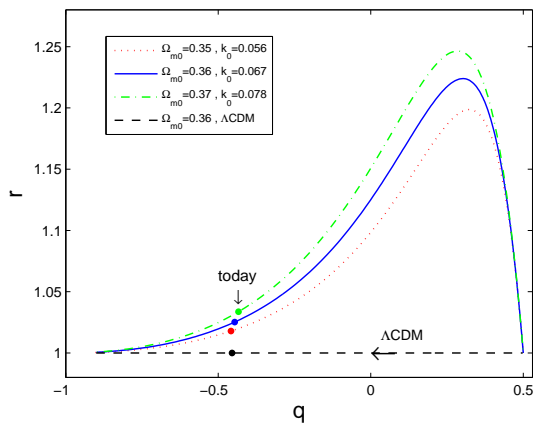}
\caption{Evolution trajectories in the $q-r$ plane. The black dashed
line represents the $\Lambda$CDM model.The today's points of purely
kinetic k-essence are also close to $\Lambda$CDM model's.}
\end{figure}

\begin{figure}
\includegraphics[width=8cm]{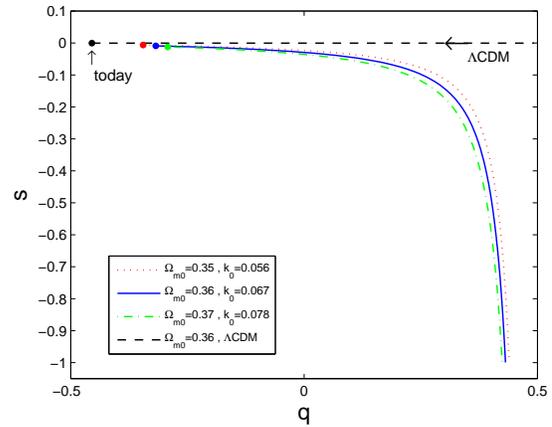}
\caption{Evolution trajectories in the $q-s$ plane.}
\end{figure}

As a complementarity, we plot the evolution trajectory in $q-r$
plane. In Fig. 2, both purely kinetic k-essence model and
$\Lambda$CDM model commence evolving from the same point in the past
$q=0.5, r=1$ which corresponds to a matter dominated SCDM (standard cold dark matter) Universe,
and end their evolution at the same point in the future. Meanwhile,
we plot the evolution trajectory in $q-s$ plane (see Fig. 3). At the beginning,
the difference between purely kinetic k-essence model and
$\Lambda$CDM model is very obvious in $q-s$ plane. When evolving,
purely kinetic k-essence model is getting closer and closer
$\Lambda$CDM model.

According the discussions above, we see that statefinder $\{r,s \}$
is fail to distinguish purely kinetic k-essence model from
$\Lambda$CDM model. Now, we apply another diagnosis method, $Om(x)$, to
distinguish them. $Om(x)$ diagnostics is defined as \cite{V2}
\begin{eqnarray}
Om(x)\equiv\frac{h^2(x)-1}{x^{3}-1},
\end{eqnarray}
where $x=1+z$ and $h(x)=H(x)/H_{0}$. The slope of $Om(x)$ can
distinguish dynamical dark energy from the cosmological constant in
a robust manner both with and without reference to the value of the
matter density \cite{V2, Feng, Granda,Zhang}. For $\Lambda$CDM, the
$Om(x)$ is
\begin{eqnarray}
Om(x)=\Omega_{\rm m0}.
\end{eqnarray}
For purely kinetic k-essence model, the $Om(x)$ is
\begin{eqnarray}
Om(x)=\frac{\Omega_{\rm
k0}\left(\frac{2k_{0}^2x^6+1}{2k_{0}^2+1}\right)^{1/2}+\Omega_{\rm
m0}x^3-1}{x^3-1}.
\end{eqnarray}

Similarly, we plot the evolutions of $Om(x)$ in Fig. 4.

\begin{figure}
\includegraphics[width=8cm]{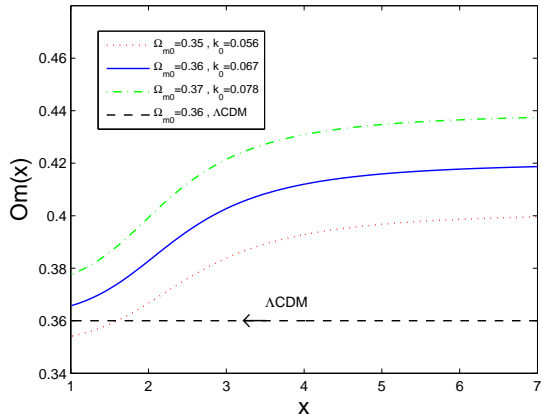}
\caption{The $Om(x)$ diagnostic}
\end{figure}

For purely kinetic k-essence model, we plot evolution trajectories
of $Om(x)$ with the values \cite{Yang09} constrained from 307 SNIa
data \cite{Kowalski}, the shift parameter $R$ \cite{wang}, and the
acoustic scale $l_{\rm a}$ \cite{wang}. At present, $Om(x)$ of
purely kinetic k-essence model is greater than that of $\Lambda$CDM
model when $\Omega_{\rm m0}=0.36$ and $\Omega_{\rm m0}=0.37$, while
less than that of $\Lambda$CDM model when $\Omega_{\rm m0}=0.35$.
However, the difference between purely kinetic k-essence model and
$\Lambda$CDM model isn't obvious near present. It is obviously that
the deviations of $Om(x)$ between purely kinetic k-essence and
$\Lambda$CDM is less than $0.06$ ($\Delta Om(x)<0.06$) even at
$z\leq 6$. Namely, the $Om(x)$ cannot discriminate those two models
at $68.3\%$ confidence level.

To understander the results above more well, we calculate analytically the lowest order of $h(x)=H(x)/H_{0}$ of
purely kinetic k-essence model and $\Lambda$CDM model. For $\Lambda$CDM model, we find
$h(x)\simeq 1+\frac{3}{2}z\Omega_{\rm
m0}$ at low redshifts. For purely kinetic k-essence model, we find $h(x)\simeq 1+\frac{1}{2}z\left[3\Omega_{\rm
m0}+\frac{6k_{0}^2(1-\Omega_{\rm
m0})}{2k_{0}^2+1}\right]$ at low redshifts. Taking $\Omega_{\rm m0}=0.36$ and $k_0=0.067$ \cite{Yang09}, we find the deviation of $Om(x)$ between purely kinetic k-essence and $\Lambda$CDM is very small for $z=0.01$: $\Delta Om(x)\simeq 0.0001$. So the $Om(x)$ cannot discriminate these two models at low redshifts.

\section{Conclusions and discussions}
In this Letter, we applied two geometrical diagnostics of dark
energy, involving the statefinder $\{r,s\}$ and $Om(x)$, to
distinguish purely kinetic k-essence with Dirac-Born-Infeld-like Lagrangian from $\Lambda$CDM model. We
plotted the evolution trajectories in statefinder $s-r$ plane. We found that the current values of purely kinetic k-essence are
close to the $\Lambda$CDM fixed point. The `distant' between two
models cannot be identified explicitly. Obviously, the distances
between these cases can't be easily measured. Therefore, the
statefinder cannot differentiate purely kinetic k-essence model from
$\Lambda$CDM model at $68.3\%$ confidence level for $z\ll 1$. As
another diagnostic method, $Om(x)$ is widely used to distinguish
different dark energy models. We found, however, the $Om(x)$ also
cannot discriminate purely kinetic k-essence model and $\Lambda$CDM
model at $68.3\%$ confidence level. In order to differentiate these
two models, it is necessary to find a new method.

\begin{acknowledgments}
This study is supported in part by Research Fund for Doctoral
Programs of Heibei University No. 2009-155, and by Open Research
Topics Fund of Key Laboratory of Particle Astrophysics, Institute of
High Energy Physics, Chinese Academy of Sciences, No.
0529410T41-200901.
\end{acknowledgments}

\end{document}